\documentclass[reprint,
               superscriptaddress,
               nobibnotes,
               amsmath,
               amssymb,
               aps,
               hidelinks, 
               pre]{revtex4-2}
\usepackage{showyourwork}
\usepackage[version=4]{mhchem}
\usepackage{graphicx}
\usepackage{bm}
\usepackage{siunitx}
\usepackage{diagbox}
\usepackage{xfrac}
\DeclareSIUnit{\angstrom}{\text{Å}}
\newcommand{\pr}[2]{p(#1\mid#2)}
\newcommand{\papertitle}{Lost in Translation: Simulation-Informed Bayesian Inference Improves Characterisation of Molecular Motion From Neutron Scattering}
\usepackage[capitalise]{cleveref}
\creflabelformat{equation}{#2\textup{#1}#3}
\makeatletter
\renewcommand{\frontmatter@abstractheading}{%
  \vspace{0.5pc}
  \centerline{\bfseries\MakeUppercase\abstractname}%
  \vspace{0.5pc}%
}
\makeatother

\makeatletter
\def\maketitle{
\@author@finish
\title@column\titleblock@produce
\suppressfloats[t]}
\makeatother

\begin{document}

\let\oldaddcontentsline\addcontentsline
\renewcommand{\addcontentsline}[3]{}

\title{\papertitle}

\author{Harry Richardson}
  \affiliation{Centre for Computational Chemistry, School of Chemistry, University of Bristol, Cantock's Close, Bristol, BS8 1TS, UK.}
\author{Kit McColl}
  \affiliation{Department of Materials, University of Oxford, Parks Road, Oxford OX1 3PH, UK.}
  \affiliation{The Faraday Institution, Quad One, Harwell Science and Innovation Campus, Didcot, OX11 0RA, UK.}
\author{Gøran J. Nilsen}
  \affiliation{ISIS Pulsed Neutron and Muon Facility, Rutherford Appleton Laboratory, Didcot, OX11 0QX, UK.}
  \affiliation{Department of Mathematics and Physics, University of Stavanger, 4036 Stavanger, NO.}
\author{Jeff Armstrong}
\email{jeff.armstrong@stfc.ac.uk}
  \affiliation{ISIS Pulsed Neutron and Muon Facility, Rutherford Appleton Laboratory, Didcot, OX11 0QX, UK.}
  \affiliation{Department of Chemistry, University of Bath, Claverton Down, Bath, BA2 7AY, UK.}
\author{Andrew R. McCluskey}
\email{andrew.mccluskey@bristol.ac.uk}
  \affiliation{Centre for Computational Chemistry, School of Chemistry, University of Bristol, Cantock's Close, Bristol, BS8 1TS, UK.}
  \affiliation{Diamond Light Source, Harwell Campus, Didcot, OX11 0DE, UK.}

\begin{abstract}
Quasi-elastic neutron scattering (QENS) probes atomic and molecular motion on length and time scales central to catalysis, energy materials, and gas adsorption. 
However, conventional analytical fitting of QENS spectra often fails to uniquely determine the underlying dynamics.
The flexibility of simplified line-shape models can make spectra generated by distinct physical processes statistically indistinguishable, leading to ambiguous or inaccurate mechanistic interpretation.
By integrating molecular dynamics simulations, physically derived $Q$-dependent scattering models, Bayesian model discrimination, and polarisation analysis, we demonstrate that QENS can, for the first time, resolve anisotropic rotational motion in liquid benzene, a prototypical aromatic molecule relevant to microporous catalysis. 
The extracted spinning and tumbling diffusion coefficients suggest stronger anisotropy than previously recognised.
This integrated, Bayesian evidence-based analytical framework defines a new paradigm for QENS, enabling direct resolution of the rotational and translational dynamics that govern molecular interactions and transport; the fundamental processes and rate-limiting steps in confined hydrocarbon catalysis.
\end{abstract}

\maketitle

% \section{Introduction}
% \label{sec:intro}

% QENS in chemistry, specifically QENS for rotational dynamics, with some choice examples. 
Progress in catalysis and materials chemistry relies on resolving molecular motion under confinement and in complex environments. 
Quasi-elastic neutron scattering (QENS), owing to its exceptional sensitivity to hydrogen, directly probes these processes over angstrom to nanometre length scales and sub-picosecond to nanosecond timescales, precisely matching the spatial and temporal regimes that govern bonding interactions and molecular transport in catalytic and functional materials~\cite{OMalley2016, Han2024, Jobic2000}.
Confined translational and rotational diffusion is often the key bottleneck in the catalytic efficiency of these materials. 
Accurate and atomistically resolved descriptions of these molecular motions, therefore, drive the design principles for higher-performing catalysts~\cite{Kolokolov2017}.
For example, high rotation rates in catalytic pores have been associated with reduced catalytic efficiency, as the reactants cannot translate easily through narrow openings~\cite{Dunn2024}
More generally, QENS is widely applied across the physical sciences, where it can probe ionic migration~\cite{Schwaighofer2025}, elucidate the diffusion of water through microporous polymer membranes~\cite{Wang2024}, and study protein dynamics in solution~\cite{Sarter2024}. 

% What does QENS measure (mentioning incoherent/coherent)
In principle, QENS enables the quantification of molecular rotation and translation by measuring the total dynamic structure factor, $S(Q,\omega)$, as a function of wavevector, $Q$ and energy transfer, $\omega$, which fully encodes the spatial and temporal atomic correlations of the system.
The total dynamic structure factor is the sum of coherent and incoherent components; the incoherent part, $S_{\mathrm{inc}}(Q, \omega)$, describes only the self-dynamics of atoms, while the coherent signal also contains information about collective dynamics~\cite{Sarter2024, Burankova2014}. 
As is the case with other scattering methods, the inversion of the dynamic structure factor from the unintuitive momentum and frequency spaces to a more relatable spatial and temporal description is often not straightforward ~\cite {Morton2025}. 
The system's dynamics are encoded in the line shape of the $\omega$-dependent QENS spectra, which is convoluted with the instrument-specific spectral resolution.
The evolution of the line shape with $Q$ then describes the contributions from the system's rotational and translational processes. 

% Current QENS analysis isn't great. 
Typically, QENS spectra are analysed by fitting a selected number of Lorentzian functions independently at each $Q$. The extracted linewidths and amplitudes are then interpreted in terms of their $Q$-dependence to infer the underlying rotational and translational mechanisms.
This approach yields a high-dimensional, poorly constrained parameter space that can easily lead to ``statistical bottlenecks'' and degenerate descriptions of the same data~\cite{Mamontov2019}, which can be fitted by multiple conflicting models, thereby undermining the physical reliability of the results. 
Furthermore, QENS analysis often relies on oversimplified models that potentially distort the dynamical behaviour of the experimental system~\cite{Hernandez2021}, making it challenging and subjective to draw physically meaningful interpretations.
Finally, in the analysis of hydrogenous systems, it is typically presumed that the total dynamic structure factor consists of purely incoherent scattering, due to the large incoherent scattering cross section of \ce{^1H}. 
However, the development of QENS with polarisation analysis (p-QENS) has shown that coherent contributions can remain significant, and neglecting them can compromise the inferred dynamics~\cite{Sarter2024}.

% Problem Statement and our solution
As catalytic and functional materials and processes grow in complexity, rigorous discrimination between competing transport mechanisms becomes essential. 
Addressing model degeneracy, global $Q$-constraints, and coherent-scattering effects within a unified framework is therefore necessary to obtain physically reliable insight from QENS measurements.
Previous work has attempted to address these issues individually~\cite{Arbe2020, Jobic2000, Sivia1992, Voneshen2017}, but no prior approach has addressed them simultaneously. 
In this work, we overcome these challenges by showing how it is possible to (i) develop and use accurate analysis models, based on the physics and chemistry of our system; (ii) reduce the number of fitting parameters by constraining the fit as a function of $Q$; (iii) empirically compare different $Q$-dependent models, using Bayesian inference; and (iv) address the presence of coherent scattering by taking advantage of p-QENS.
To implement this framework, we use insights from molecular dynamics (MD) simulations to guide the interpretation of QENS spectra of liquid benzene.
Benzene is used as a test case because, despite its structural simplicity, its complex dynamics, including multiple rotational modes~\cite{Gillen1972, Rothschild2012}, remain poorly characterised.
Our approach is demonstrated through a state-of-the-art study in which the complex dynamics, and the rotational modes of liquid benzene, are quantified. 
Predictions from molecular dynamics simulations are used to inform experimental design and guide QENS analysis. 
Finally, we show that this comprehensive modelling-theory combination can extract previously unobtained information about the dynamics in this classic, well-studied system~\cite{Kikuchi2023}.

% \section{Results and Discussion}
% \label{sec:res}

\subsection*{Simulations as a Qualitative Model}
\label{sec:sims}

% Simulations overlap but aren't perfect, hence qualitative
The time and length scales covered by QENS perfectly overlap with those typically probed by classical MD simulation~\cite{Armstrong2022}. 
This correspondence makes MD simulations a powerful complementary tool for interpreting QENS data, enabling the computation of a ``noise-free'' dynamic structure factor directly from the simulation trajectory~\cite{Kneller1995, Rog2003, Hinsen2012, Goret2017}. 
A perfectly accurate MD simulation could be directly compared with observed QENS measurements to provide atomistic insights that are not obtainable from experiments alone. 
However, the interatomic potentials used in classical MD simulations approximate atomic and molecular interactions and, therefore, cannot achieve chemical accuracy. 
Nevertheless, classical simulations can still be used to inform the analysis of QENS measurements in a range of different ways~\cite{Morton2025, Petersen2022}, including the use of real-space analysis of molecular dynamics simulations as a qualitative model of the dynamics.

\begin{figure}
    \centering
    \includegraphics[width=\columnwidth]{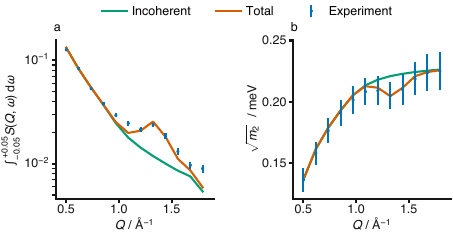}
    \vspace{-1.5\baselineskip}
    \caption{Comparison of the structure and dynamics of experimental (blue, error bars indicate a single standard deviation) and simulated (total dynamic structure factors, orange and incoherent-only, green) benzene: (a) a \qty{\pm0.05}{\milli\electronvolt} peak integral of the QENS spectra and (b) the reduced second spectral moment.}
    \label{fig:structure}
    \vspace{-1\baselineskip}
    \script{struc_prod.py}
\end{figure}

% Simulation comparison with experiment, failure of the incoherent assumption
The incoherent and total dynamic structure factors were calculated from MD simulations of benzene at \qty{290}{\kelvin}, using MDANSE~\cite{Goret2017} (full simulation details available in \ref{sec:mdsim}). 
\cref{fig:structure} compares the experimental QENS data (collected at the IRIS instrument, ISIS Neutron and Muon Source~\cite{Carlile1992}, using the PG 002 analyser crystal, giving dynamic ranges for $Q$ of $\qtyrange{0.5}{1.8}{\per\angstrom}$ and $\omega$ of \qty{\pm 0.4}{\milli\electronvolt}, see \ref{sec:qens} for further experimental details).
We present the incoherent-only and total dynamic structure factors, as characterised by the elastic peak integral and the reduced second spectral moment, respectively, as descriptors of liquid structure and dynamics~\cite{Armstrong2016}. 
Although the simulations generally capture the structure and dynamics, there is a discrepancy between the experimental and incoherent-only simulation data.
This feature is on the length scale of a single benzene molecule, i.e., $Q\approx\qty{1.3}{\per\angstrom}$ corresponding to around $\qty{4.8}{\angstrom}$, and is indicative of de Gennes narrowing (see \cref{fig:si_degenne}), which has been observed previously in the coherent dynamic structure factor~\cite{Arbe2023, Wu2018}. 
This anomaly indicates that even for hydrogenous benzene, the common approach to overlook coherent scattering is not suitable~\cite{Sarter2024}, necessitating either more complex fitting models that represent coherent scattering, or making use of p-QENS, which allows the coherent and incoherent dynamic structure factors to be separated~\cite{Nilsen2017}.
%

\begin{figure}
    \centering
    \includegraphics[width=\columnwidth]{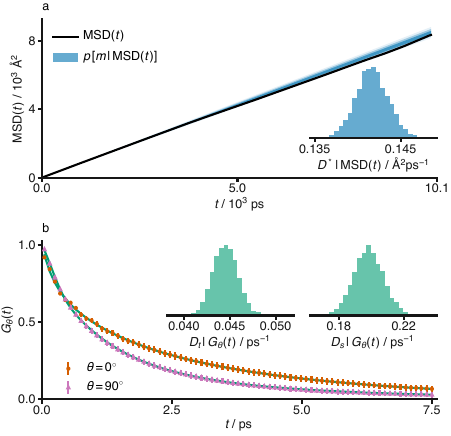}
    \vspace{-1.5\baselineskip}
    \caption{Real-space analysis of dynamics from simulation; (a) the observed mean-squared displacement (black line) with posterior distribution of linear models (blue shading indicating $1\sigma$, $2\sigma$, and $3\sigma$ credible intervals), (a, inset) the marginal posterior distribution of self-diffusion coefficients from the linear models in (a), (b) the rotational autocorrelation function of vectors parallel ($\theta=\qty{0}{\degree}$, orange dots with error bars indicating a single standard deviation) and perpendicular ($\theta=\qty{90}{\degree}$, pink triangles) to the benzene principal axis of rotation, and maximum \emph{a posteriori} fitted models (solid green lines), and (b, inset) the marginal posterior distributions of the rotational diffusion coefficients found by sampling Eqs.~\ref{eqn:exp} \& \ref{eqn:exp2} given both rotational autocorrelation functions.}
    \label{fig:simulation}
    \vspace{-1\baselineskip}
    \script{four_panel_sim.py}
\end{figure}

% Real-space analysis (simulation)
To choose the appropriate analytical model for interpreting the QENS data, we can analyse components of the molecular dynamics simulations, which provide a real-space description of molecular motion. 
First, we consider translational motion, characterised by the mean-squared displacement (\cref{fig:simulation}a), which shows a single linear regime, indicating the presence of a single translational mode with a self-diffusion coefficient of %
  \qty{0.142 \pm 0.002}{\angstrom\squared\per\pico\second}\unskip\unskip\label{output/kinisi_D.txt}\unskip%
~\footnote{Here, the uncertainty is a single standard deviation of the marginal posterior distribution.} (see \ref{sec:kinisi} for details).
While benzene has been shown to hop between transient cages~\cite{Magro2005}, these hops are too short-lived to be observed by QENS measurements. 
In the QENS data, a single translational mode would present itself as an $\omega$-dependent Lorentzian function, the width of which will increase linearly as a function of $Q^2$, i.e., the Fickian diffusion QENS model~\cite{Jobic2000}. 
%

% anisotropy of benzene in real space
Next, we consider rotational motion from the MD simulations. 
The benzene molecule is an oblate symmetric top with $D_{\textrm {6h}}$ symmetry.
It has been observed from model-fitting to MD simulation, NMR and Raman spectroscopy data that the spinning of the benzene molecule around its principal axis is \num{2.3\pm0.9}~\footnote{The uncertainty here indicates a single standard deviation from 17 samples.} times faster than the tumbling through the plane of the molecule~\cite{Schwartz2005}. 
However, previous QENS measurements were unable to discriminate between the spinning and tumbling motions~\cite{Kikuchi2023}.

To probe the rotational anisotropy in our simulation, the rotational autocorrelation functions for vectors parallel and perpendicular to the principal axis of rotation were calculated (\cref{fig:simulation}b). 
The perpendicular rotational autocorrelation function decays significantly faster than the parallel one, as the former is associated with both fast spinning and tumbling. 
In contrast, the latter is influenced by tumbling only. 
For an anisotropic axially symmetric rotor, like benzene, as the autocorrelation time, $t$, varies, the perpendicular rotational autocorrelation function will follow, 
\begin{equation}
    G_{\qty{90}{\degree}}(t) = \frac{1}{4}\exp{\left(-6D_{t}t\right)} + \frac{3}{4}\exp{\left[-\left(2D_{t} + 4 D_{s}\right)t\right]}, 
    \label{eqn:exp}
\end{equation}
and the parallel autocorrelation time is described by, 
\begin{equation}
    G_{\qty{0}{\degree}}(t) = \exp{\left(-6D_tt\right)},
    \label{eqn:exp2}
\end{equation}
where $D_t$ and $D_s$ are the tumbling and spinning diffusion coefficients, respectively (see \ref{sec:derive} for the full derivation). 
We use these models, along with additional functional descriptions of fast vibrations and librations (full details of this analysis are given in \ref{sec:rotational}), to simultaneously estimate the posterior distributions for $D_t$ and $D_s$ (\cref{fig:simulation}b, inset).
The difference between these two diffusion coefficients is around $%
  \qty{0.15}{\per\pico\second}\unskip\unskip\label{output/D_rot_diff.txt}\unskip%
$, which is in principle within the temporal resolution of many QENS instruments. 
Notably, we find a larger anisotropy, of around %
  \num{4.5}\unskip\unskip\label{output/D_rot_ratio.txt}\unskip%
, than in previous experimental and simulation studies. 
However, the model-fitting approach used previously is inconsistent: a single exponential function is used to infer a single effective timescale; this fails to model higher-energy local motions and artificially increases $D_t$ (discussed in detail in~\ref{sec:previous}), leading to an underestimation of the rotational anisotropy of benzene. 

% Outline the QENS model
To assess whether the rotational anisotropy of benzene can be observed in QENS measurements, an appropriate analytical model for the dynamic structure factor must be used. 
To our knowledge, no such model has been applied previously; hence, it is necessary to derive a $Q$/$\omega$-dependent analytical model for the anisotropic rotation of an axially symmetric rotor. 
This derivation is given in \ref{sec:derive}, and when expanded to the 2nd order spherical Bessel functions is,
\begin{equation}
\begin{aligned}
S_R(Q, \omega) = & \,
j_0^2(QR)\,\delta(\omega) + 3\,j_1^2(QR)
\left[
\frac{1}{\pi}
\frac{\Gamma_1}{\omega^2 + \Gamma_1^2}
\right] \\ 
 & + 5\,j_2^2(QR)
\left[
\frac{1}{4}
\frac{1}{\pi}
\frac{\Gamma_2}{\omega^2 + \Gamma_2^2} 
+
\frac{3}{4}
\frac{1}{\pi}
\frac{\Gamma_3}{\omega^2 + \Gamma_3^2}
\right],
\label{eqn:fit}
\end{aligned}
\end{equation}
where $j_n(Q,R)$ represents the $n$-th order spherical Bessel function with a radius of gyration $R$, which was fixed to \qty{2.48}{\angstrom} for benzene and $\delta(\omega)$ is a Dirac delta function. 
The individual $\Gamma$ values are made up from the spinning and tumbling rotational diffusion coefficients as follows: $\Gamma_1 = D_s + D_t$, $\Gamma_2 = 6D_t$, and $\Gamma_3 = 4D_s + 2D_t$.

\subsection*{Assessing the Anisotropic Model For Simulated QENS Data}
\label{sec:dynamic}

% Introductory paragraph contextualise the aim. 
The real-space analysis of the simulation suggests that QENS may be used to distinguish between the two rotational modes of benzene.
If the rates of the two rotations are too similar to discern with QENS, there is no advantage to the more complex anisotropic model in \cref{eqn:fit}, over the simpler isotropic rotation model (\cref{eqn:iso})~\cite{Bee1988}. 
To determine which model better represents the observed data, we turn to Bayesian model selection, which allows models with different numbers of fitted parameters to be compared, penalising more complex models when they are unnecessary~\cite{Sivia1992, Sivia2006, McCluskey2020}. 

% Brief introduction to Bayesian model selection
Bayesian model selection involves estimating the Bayesian evidence for some model, $m$, 
\begin{equation}
    \pr{m}{\mathbf{D}} \propto \pr{\mathbf{D}}{m}p(m),
\end{equation}
where $p(m)$ is a prior probability associated with the model and $\pr{\mathbf{D}}{m}$ is the likelihood that the data, $\mathbf{D}$, would be observed given the model, regardless of the model parameters, $\bm{\theta}$. 
This likelihood is found as the integral over the posterior parameter volume,
\begin{equation}
    \pr{\mathbf{D}}{m} = \idotsint_{\bm{\theta}_m} \pr{\mathbf{D}}{\bm{\theta}_m,\,m}\pr{\bm{\theta}_m}{m}\;\mathrm{d}\bm{\theta}_m.
    \label{eqn:inte}
\end{equation}
Each additional parameter in an analytical model increases the dimensionality of the integral, leading to an exponential increase in the effective prior volume. 
Therefore, if the added parameter is not informative, the region of high likelihood occupies a much smaller fraction of that volume, thereby reducing the integral and, consequently, the Bayesian evidence.
For there to be ``strong evidence'' for a more complex model, $m_1$, (i.e., the anisotropic model, with more parameters) the natural logarithm of this model's evidence should be at least \num{5} greater than for the simpler model, $m_0$, given the same data, i.e., $\ln\left[\pr{m_1}{\mathbf{D}}\right] \geq \ln\left[\pr{m_0}{\mathbf{D}}\right] + 5$~\cite{Kass1995}. 

% Decomposed simulation shows anisotropy in q-, omega-space
% Compare with isotropic model (simpler)
To assess whether our QENS measurement can distinguish between the two rotational modes, we first consider a simulated dynamic structure factor that contains only rotational information. 
We achieve this by centring each benzene molecule in the calculated simulation trajectory at its initial molecular centre of mass at each simulation timestep, thereby removing translational motion from the trajectory (details are provided in \ref{sec:decomp}).
From this trajectory, we calculate the incoherent dynamic structure factor with a symmetric $\omega$-dynamic range of $\qty{\pm0.4}{\milli\electronvolt}$, matching that of the experimental data collected at IRIS and assuming Poisson counting statistics to estimate the uncertainty. 
Given this simulated signal, the Bayesian evidence for each analytical model was estimated using nested sampling~\cite{Skilling2004}, with the prior probability for each model set to \num{1}. 
Comparing these evidences showed strong support for the more complex anisotropic model, with a log-Bayesian evidence of $%
  $-56.0\unskip$\unskip\label{output/aniso_rot_only_evidence.txt}\unskip%
$, compared to $%
  $-78.2\unskip$\unskip\label{output/iso_rot_only_evidence.txt}\unskip%
$ for the isotropic model. 

\begin{figure}
    \centering
    \includegraphics[width=0.67\columnwidth]{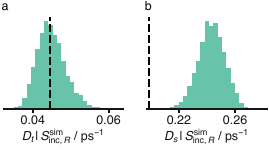}
    \vspace{-0.5\baselineskip}
    \caption{Marginal posterior distributions of the (a) tumbling and (b) spinning diffusion coefficients for \cref{eqn:fit} for the incoherent dynamic structure factor computed from the rotation-only simulation, vertical black lines show the mean of the marginal posterior distributions from the rotational autocorrelation function.}
    \label{fig:rotational}
    \vspace{-1\baselineskip}
    \script{rot_sim_ds.py}
\end{figure}

% Posterior from the pure rotation
Nested sampling simultaneously estimates the posterior distribution of the parameters; the marginal posterior distributions for the two rotational diffusion coefficients are shown in \cref{fig:rotational}.
These distributions are compared with the mean values of the rotational diffusion coefficients obtained directly from the simulation's rotational autocorrelation function. 
The posterior distribution of \cref{eqn:fit} gives a larger value for the rotational diffusion coefficient of spinning when compared with those from the rotational autocorrelation function, likely arising from the difficulty of empirically separating motions within the incoherent dynamic structure factor, even for the deconvolved rotational model.

%

% Total simulation needs high energy dynamic range to show anisotropy (LOST IN TRANSLATION!)
% Also suppresses the perpendicular rotational diffusion
Having shown that, in the hypothetical case where only rotation is present in the incoherent dynamic structure factor, one can observe individual rotations, we now consider whether the anisotropic model retains strong support when translational motion is also present. 
Therefore, we convolve our model for rotation with one for Fickian diffusion at each $Q$-value, 
\begin{equation}
    S_{\textrm{inc}}(Q, \omega) = e^{-Q^2 \langle u^2 \rangle / 3} \left[\frac{1}{\pi} \frac{D^*Q^2}{\omega^2 + (D^*Q^2)^2} \right] \otimes S_{\mathrm{R}}(Q, \omega),
    \label{eqn:full}
\end{equation}
where $D^*$ is the Fickian translational self-diffusion coefficient, $\langle u^2 \rangle$ is the mean square displacement for the Debye-Waller factor, and $S_{\mathrm{R}}(Q, \omega)$ is either \cref{eqn:fit} or the isotropic rotational model (\cref{eqn:iso}). 
\cref{eqn:full} is significantly more constrained, using either the isotropic or anisotropic models, than the traditional approach to QENS analysis outlined in the introduction would achieve. 

\begin{figure}
    \centering
    \includegraphics[width=\columnwidth]{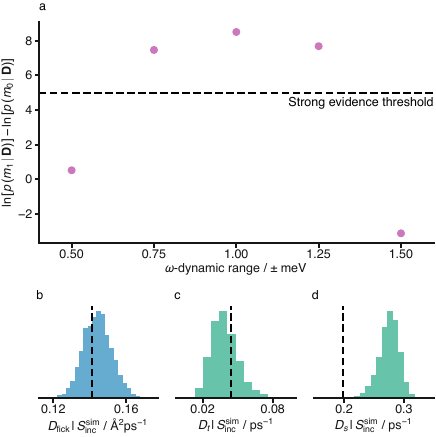}
    \vspace{-1.5\baselineskip}
    \caption{Analysis of $\omega$-dynamic range dependence of Bayesian evidence using QENS from the full simulation: (a) the Bayesian evidence for the anisotropic model compared with the isotropic model as a function of $\omega$-dynamic range, where the threshold for strong evidence is indicated with the dashed black line, and (b) the Fickian self-diffusion coefficient, (c) tumbling, and (d) spinning rotational diffusion coefficients marginal posterior distributions, obtained for the anisotropic model, given the simulated incoherent QENS signal at an $\omega$-dynamic range of $\qty{\pm1.25}{\milli\electronvolt}$, compared with the mean of the distributions obtained directly from simulation trajectory (dashed vertical black lines).}
    \label{fig:energy}
    \vspace{-1\baselineskip}
    \script{evidence_full_sim.py}
\end{figure}

\cref{fig:energy}a presents the difference in Bayesian evidence between the isotropic and anisotropic models, using the incoherent dynamic structure factor from the full simulation trajectory. 
At a small $\omega$-dynamic range of $\qty{\pm0.5}{\milli\electronvolt}$, slightly above that of the IRIS measurements, it is not possible to distinguish between the two models. 
However, as the $\omega$-dynamic range is increased, the Bayesian evidence for the more complex model increases, and from $\qtyrange{\pm 0.75}{\pm1.25}{\milli\electronvolt}$ there is strong evidence for the anisotropic model. 
Representative marginal posterior distributions of the translational self-diffusion coefficient and rotational diffusion coefficients are given in Figs.~\ref{fig:energy}b, \ref{fig:energy}c, and \ref{fig:energy}d at \qty{\pm1.25}{\milli\electronvolt}.
While the two rotational diffusion coefficients are distinct, their values differ slightly from those found by analysing the decomposed simulation.
This difference may be due to the assumption that translation and rotation are completely uncorrelated, which is not accounted for when the molecular translation is removed, or the use of the numerical convolution to evaluate \cref{eqn:full}. 
At larger $\omega$-dynamic ranges, high-energy local motions that were observed in the real-space analysis above, are present in the simulated QENS signal, this was also characterised experimentally previously \cite{Kikuchi2023}. 
Neither \cref{eqn:full} nor the isotropic model describe these high-energy motions, therefore, the Bayesian evidence suggests that we should accept the ``null hypothesis'' of the less complex model. 

\subsection*{Application to Experimental QENS Data}
\label{sec:exp}

\begin{figure*}
    \centering
    \includegraphics[width=0.8\textwidth]{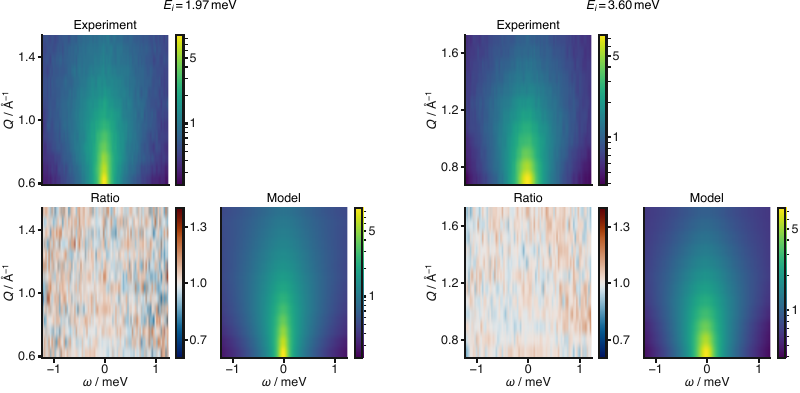}
    \vspace{-0.5\baselineskip}
    \caption{Anisotropic model maximum \emph{a posteriori} given experimental LET data at incident energies of $\qtylist{1.97;3.60}{\milli\electronvolt}$: comparison of the incoherent dynamic structure factor from the experimental data, co-refined anisotropic model, and their ratio. Individual $Q$-dependent line fits available in Figs.~\ref{fig:si_model360_fit}~\&~\ref{fig:si_model197_fit}.}
    \label{fig:qens}
    \vspace{-1\baselineskip}
    \script{heatmap.py}
\end{figure*}

% Throw it at the experiment, what comes out?
We have shown that to observe rotational anisotropy in benzene from a simulated QENS signal, it is necessary to have an $\omega$-dynamic range of $\qtyrange{\pm0.75}{\pm1.25}{\milli\electronvolt}$ and to minimise the presence of coherent scattering. 
This dynamic range exceeds that of the IRIS instrument with the PG 002 analyser crystal, and, as shown in \cref{fig:structure}, it is not possible to remove coherent scattering from this signal.
Therefore, we measured the QENS at the LET instrument (ISIS Neutron and Muon Source) in the polarisation analysis configuration using incident energies of \qtylist{1.97;3.60}{\milli\electronvolt}, this data was rebinned and truncated to give an $\omega$-dynamic range of $\qty{\pm1.25}{\milli\electronvolt}$~\cite{Nilsen2017}.
We chose the largest $\omega$-dynamic range to account for the possibility of faster spinning that was observed in simulation. 
Polarisation analysis allows for the decomposition of the total dynamic structure factor into its individual components; this means that the Bayesian evidence for the two analysis models can be found given the purely incoherent data. 
There was strong support for the more complex anisotropic model, with a log-Bayesian evidence of %
  $-2874.3\unskip$\unskip\label{output/evidence_aniso_let.txt}\unskip%
, compared to %
  $-3032.8\unskip$\unskip\label{output/evidence_iso_let.txt}\unskip%
 for the isotropic model, given the noisy experimental data. 
\cref{fig:qens} compares the $Q$-dependent anisotropic model with the measured LET data at each incident energy, showing strong agreement between the two (this is further compared with the lower-evidence isotropic model in Figs. \ref{fig:si_model360_fit} \& \ref{fig:si_model197_fit}); the ratio between the model and the experiment shows noise, with no structuring present in the error (see~\ref{sec:model_fit} for analysis details). 
%

\begin{figure}
    \centering
    \includegraphics[width=\columnwidth]{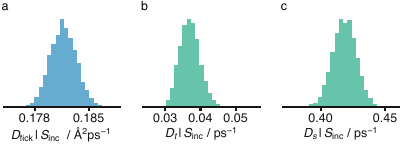}
    \vspace{-1.5\baselineskip}
    \caption{The marginal posterior distributions for the anisotropic model given data at both incident energies; (a) the Fickian self-diffusion coefficient, and (b) the tumbling and (c) spinning rotational diffusion coefficients.}
    \label{fig:hist}
    \vspace{-1\baselineskip}
    \script{hists.py}
\end{figure}

% Results from the LET analysis
Given the LET data, the full posterior distribution for the anisotropic analytical model was also found. 
All three experimental diffusion coefficients differ slightly from those obtained from the MD simulation, with the experiment showing faster Fickian diffusion and parallel rotational diffusion, and slower perpendicular rotational diffusion.
The marginal posterior distribution for the Fickian self-diffusion coefficient is shown in \cref{fig:hist}a, with a mean and $1\sigma$ credible interval of $%
  \qty{0.182 \pm 0.002}{\angstrom\squared\per\pico\second}\unskip\unskip\label{output/D_fick_let.txt}\unskip%
$.
This value agrees well with previous QENS analysis that found the self diffusion coefficient to be $\sim\qty{0.16}{\angstrom\squared\per\pico\second}$~\cite{Kikuchi2023} and other experimental measurements, such as the $\qty{0.186}{\angstrom\squared\per\pico\second}$ at \qty{288.2}{\kelvin}~\cite{Collings1970}. 

Significantly, the experimental estimates for the two rotational diffusion coefficients are distinct and non-zero (\cref{fig:hist}b and \cref{fig:hist}c); by addressing the issues highlighted in the introduction, it is possible, for the first time, to distinguish the anisotropy of the benzene rotation with QENS. 
Previously, it was not possible to observe anisotropy, and a single rotational diffusion coefficient was obtained from QENS data~\cite{Kikuchi2023}. 
This single value closely matches the average of the tumbling and spinning diffusion coefficients determined in this work. 
The marginal posterior distributions for $D_s$ and $D_t$ indicate faster and slower rotation, respectively, than observed in our simulations, leading to an anisotropy ratio of around %
  \num{11}\unskip\unskip\label{output/D_ratio_let.txt}\unskip%
.
This anisotropy ratio is greater than that found from simulation alone, which may be rationalised by recognising that experimentally, liquid benzene has been shown to form `T'- and `Y'-shaped perpendicular dimers, as well as parallel $\pi$-stacked agglomerates~\cite{Falkowska2016, Headen2019}; structures that are not observed in our simulations (see~\ref{sec:comp_marta}) and would make tumbling energetically unfavourable compared to spinning. 
Though interesting, this elevated anisotropy requires further investigation to conclusively say it was underestimated in previous work.
By using simulations as a quantitative model for the experimental data, it has been possible to observe anisotropy that could not be measured previously with QENS. 

% \section{Conclusions}
% \label{sec:con}

We have derived a detailed, fully $Q$-dependent analytical model for anisotropic rotational motion and explicitly evaluated its statistical sensitivity within the $Q$- and $\omega$- ranges accessible to QENS, using Bayesian analysis informed by molecular dynamics simulations. 
Furthermore, we employed Bayesian model selection to compare this model with simpler, lower-parameter descriptions, demonstrating that an anisotropic model provides a more accurate and physically nuanced account of the data. 
Importantly, analysis of the simulated $\omega$-sensitivity identified the regime in which the anisotropic model is statistically identifiable, guiding experimental design. 
Polarisation-resolved QENS then isolated the incoherent scattering component, which is directly described by the analytical model, ensuring consistency between theoretical assumptions and measured observables.
By integrating these elements, we have shown, for the first time, that QENS can fully resolve both translational and anisotropic rotational dynamics in liquid benzene, explicitly distinguishing two distinct rotational modes and suggesting that previous estimates of anisotropy in benzene may require reconsideration.
More broadly, this enhanced resolution of translational and rotational motion has significant implications for catalysis and functional porous materials. 
Quantitative characterisation of molecular diffusion and anisotropic rotation within zeolite catalysts provides direct measures of bonding constraints and local transport processes, which determine catalytic rate and conversion efficiency.
We hope that this work will inspire others to consider more carefully the information that may be drawn from their QENS data and enable other researchers to maximise the information obtained from this powerful technique. 

\section*{Author Contributions}

CRediT author statement: 
H.R.: Data curation, Formal analysis, Investigation, Methodology, Software, Visualisation, Writing - original draft.
K.M.: Conceptualization, Funding acquisition, Investigation, Supervision, Writing - review \& editing. 
G.J.N.: Data curation, Investigation, Writing - review \& editing. 
J.A.: Conceptualization, Formal analysis, Funding acquisition, Investigation, Methodology, Project administration, Supervision, Writing - review \& editing. 
A.R.M.: Conceptualisation, Formal analysis, Funding acquisition, Investigation, Methodology, Project administration, Supervision, Visualisation, Writing - original draft. \\

\section*{Supporting Information} 

Supporting information for this work is available and includes: 
\begin{itemize}
    \item Descriptions of the simulation parameters, experimental measurements and data reduction.
    \item Details of the simulation translational motion analysis. 
    \item A more in-depth discussion of the presence of de Gennes narrowing.
    \item Mathematical derivations of the rotational autocorrelation function and dynamic structure factor for an oblate symmetric top.
    \item Details of the fitting approach used for the rotational autocorrelation function analysis. 
    \item A comparison of the rotational autocorrelation function analysis used in previous literature with that used in this work. 
    \item Information about the QENS analysis from the decomposed simulation trajectory. 
    \item Deatils of the QENS model fitting presented in the main text. 
    \item Figures showing the QENS signal for each $Q$-bin compared with the maximum \emph{a posteriori} anisotropic model at the same $Q$-bin. 
    \item The position and orientation probability density function for the benzene simulation. 
\end{itemize}

\section*{Acknowledgements}

H.R. acknowledges the Engineering and Physical Sciences Research Council for DTP funding (Grant No. EP/W524414/1).
K.M. is grateful to the Faraday Institution CATMAT project (EP/S003053/1, FIRG016) for financial support. 
This work was carried out using the computational facilities of the Advanced Computing Research Centre, University of Bristol - \href{http://www.bristol.ac.uk/acrc/}{http://www.bristol.ac.uk/acrc}.
Experiments at the ISIS Pulsed Neutron and Muon Source on IRIS (RB2420517) and LET (RB2590225) were supported by a beamtime allocation from the Science and Technology Facilities Council.
The authors would like to thank Prof. Neil Allan for his insightful feedback on this work. 

\section*{Data Availability and Reproducibility Statement}

Raw experimental QENS data is available from the ISIS Neutron and Muon Source Data Journal~\cite{iris_data, plet_data}. 
The data used to generate the figures in this paper are available on Zenodo~\cite{Richardson2026} and all of the code required to recreate them can be found, as a \textsc{showyourwork} workflow in the appropriate GitHub repository~\cite{Richardson2026a}, under MIT (code) and CC BY-SA 4.0 (figures and text) licenses.
Also present in this repository is the custom Python code used to perform the analysis presented in this work, which is similarly fully computationally reproducible.  

\bibliographystyle{achemso}
\bibliography{bib}

\let\addcontentsline\oldaddcontentsline

\onecolumngrid
\clearpage 

\appendix
\renewcommand\thesection{SI.\Roman{section}}
\counterwithout{figure}{section}
\renewcommand\thefigure{SI.\arabic{figure}}
\setcounter{figure}{0}
\counterwithout{equation}{section}
\renewcommand\theequation{SI.\arabic{equation}}
\setcounter{equation}{0}
\counterwithout{table}{section}
\renewcommand\thetable{SI.\arabic{table}}
\setcounter{table}{0}
\pagenumbering{arabic} 
\renewcommand\thepage{SI.\arabic{page}}
\addtocontents{toc}{\protect\setcounter{tocdepth}{0}}

\title{Supplemental Material for ``\papertitle''}
\maketitle

\onecolumngrid
\vspace{-1.5cm}

\section{Classical Molecular Dynamics Simulation Parameters}
\label{sec:mdsim}

Molecular dynamics simulations were carried out in LAMMPS~\cite{Thompson2022}. 
Initial configurations of the bulk liquid were constructed by the $9\times 9\times 9$ replication of a single molecule (resulting in \num{729} molecules) generated using LigParGen~\cite{Dodda2017}. 
All atomic charges and intra- and intermolecular interactions were modelled using the OPLS-AA force field~\cite{Jorgensen1996}. 
Lennard-Jones and real-space Coulombic interactions were truncated at \qty{9}{\angstrom}, with long-range electrostatics evaluated using the particle-particle particle-mesh (PPPM) method~\cite{Eastwood1984}.
The system was equilibrated for \qty{20}{\pico\second} at \qty{10}{\kelvin} with a \qty{1}{\femto\second} timestep using the NVT ensemble and a Nose-Hoover thermostat~\cite{Nos1984, Hoover1985, Martyna1992}, with a temperature damping parameter of \qty{100}{\femto\second}. 
The temperature of the system was then ramped from $\qtyrange{10}{290}{\kelvin}$ at \qty{1}{atm} in the NPT ensemble using a Nose/Hoover temperature thermostat and Nose/Hoover pressure barostat over \qty{1}{\nano\second}~\cite{Martyna1994}, with the same temperature damping as previously and a pressure damping parameter of \qty{1000}{\femto\second}. 
Following the temperature ramp, the system was run at \qty{290}{\kelvin} and \qty{1}{atm} for \qty{0.5}{\nano\second} and the trajectory frame reflecting the mean box dimensions was taken, resulting in a \qty{47.7}{\angstrom} cubic simulation cell and simulation density of \qty{0.869}{\gram\per\cubic\centi\metre} (compared with an experimental density of \qty{0.879}{\gram\per\cubic\centi\metre} for liquid benzene at \qty{293.15}{\kelvin} \cite{Malek2012}).
The production trajectory was generated from a \qty{10}{\nano\second} NVT ensemble run at \qty{290}{\kelvin}, starting from the selected frame, with frames stored every \qty{0.5}{\pico\second}. 
For the rotational autocorrelation functions discussed in the main text, the same equilibration and system size were used, but a production trajectory of \qty{1}{\nano\second} sampled every \qty{0.05}{\pico\second} was generated. 
The autocorrelation functions were then calculated over 100 independent \qty{10}{\pico\second} windows within the trajectory.

\section{Quasi-Elastic Neutron Scattering}
\label{sec:qens}

\subsection{Sample and Measurement}

On IRIS, neat benzene (Sigma-Aldrich, \qty{99.8}{\percent} purity) was loaded into aluminium sample cans, with a \qty{0.25}{\milli\meter} thick annular space, in a fume hood to minimise exposure to moisture in the air, and the cans were sealed with indium wire.
The sample was loaded onto the IRIS instrument and measured for \qty{190}{\minute} at \qty{290}{\kelvin}. 
An empty aluminium can was used as a background, and a previously measured vanadium sample served as the resolution function.

On LET, the same benzene was loaded into a \qty{15}{\milli\meter} (outer diameter) aluminium can with a \qty{0.25}{\milli\meter} thick annular space. 
The can was sealed with indium wire, and then loaded into an Orange cryostat on LET, where it was measured for around \qty{8}{\hour} (\qty{480}{\minute}) at \qty{290}{\kelvin} and \qty{2}{\hour} at \qty{260}{\kelvin}. 
A total of 4 incident energies, $E_i$, were measured using repetition rate multiplication: $\qtylist{8.61;3.60;1.95;1.23}{\milli\electronvolt}$. 
The energy resolutions at these energies were \qty{300}{\micro\electronvolt}, \qty{85}{\micro\electronvolt}, \qty{45}{\micro\electronvolt} and \qty{15}{\micro\electronvolt} and the maximum accessible energy dynamic ranges were $\qty{\pm6.88}{\milli\electronvolt}$, $\qty{\pm2.88}{\milli\electronvolt}$, $\qty{\pm1.576}{\milli\electronvolt}$ and $\qty{\pm0.992}{\milli\electronvolt}$ respectively.
The two intermediate incident energies were used for the analysis, as they provide sufficient resolution and dynamic range to capture the expected dynamics in benzene.
An empty can was used as a background, and a frozen benzene sample at \qty{260}{\kelvin} was measured and used for the resolution function (see detailed discussion of data reduction below).

Polarisation analysis on LET was achieved using a polarised neutron beam, from an $m=5$ \ce{Fe}/\ce{Si} supermirror V-cavity, and the scattered polarisation was analysed with a hyperpolarised \ce{^3He} spin-filter cell. 
A current-ramped precession coil $\pi$-flipper is placed after the polariser to measure the spin-flip and non-spin flip intensities, $S_{\mathrm{sf}}$ and $S_{\mathrm{nsf}}$, as a function of scattering angle, $\theta$, and energy transfer, $\omega$.
Measurements of the time-dependent polarisation efficiency of the instrument taken using a monitor in the direct beam were used to correct the raw spin-flip and non-spin-flip intensities according to the procedure described in \cite{Arbe2020, Cassella2019}, following which they were converted to dynamic structure factors, $S_{\mathrm{sf}}(Q, \omega)$ and $S_{\mathrm{nsf}}(Q, \omega)$. 
The spin-incoherent dynamic structure factor $S_{\mathrm{inc}}$ was finally obtained from the corrected spin flip dynamic structure factor according to:
\begin{equation}
    S_{\mathrm{inc}}(Q, \omega) = \frac{3}{2}S_{\mathrm{sf}}(Q, \omega).
\end{equation}

\subsection{Data Reduction}
\label{sec:data-red}

The experimental IRIS data was binned into \num{12} $Q$-bins, each with a width of ~\qty{0.113}{\per\angstrom} across the 51 detectors with a total $Q$-range of $\qtyrange{0.44}{1.79}{\per\angstrom}$.
A vanadium sample was used to determine the instrument's resolution function. 
The same sample was used to correct for detector efficiency, by normalising each QENS spectrum with the energy-integrated vanadium sample intensity at the corresponding $Q$-value. 
Data measured from an empty sample can were used to quantify the instrument background and were subtracted from both the benzene and vanadium data. 

The experimental p-LET data were initially binned into \num{30} $Q$-bins for each incident energy.
This data was then sampled in the chosen $Q$ range, based on the region of interest and instrumental constraints $\qtyrange{0.67}{1.72}{\per\angstrom}$, for $E_i = \qty{3.60}{\milli\electronvolt}$ and between $\qtyrange{0.59}{1.53}{\per\angstrom}$ for $E_i=\qty{1.97}{\milli\electronvolt}$. 
This resulted in 12 and 14 bins of width ~\qty{0.096}{\per\angstrom} and ~\qty{0.073}{\per\angstrom} for \qty{3.6}{\milli\electronvolt} and \qty{1.97}{\milli\electronvolt}, respectively, within the chosen $Q$-range.
Variations in instrumental detector efficiency were corrected using a “white-beam” vanadium measurement collected at the beginning of the cycle.
A frozen benzene measurement at \qty{260}{\kelvin} was used as an approximation of the instrumental resolution function. 
This frozen sample is not a perfect representation of the true resolution, as some molecular mobility remains at this temperature. 
Although vanadium is often used to estimate instrumental resolution, its scattering profile differs from that of the sample, particularly in the high-energy-transfer tails, which can introduce systematic distortions in the convolution during the final model.

\section{Evidence of de Gennes Narrowing}
\label{sec:degennes}

\begin{figure}[h]
    \centering
    \includegraphics[width=0.4\columnwidth]{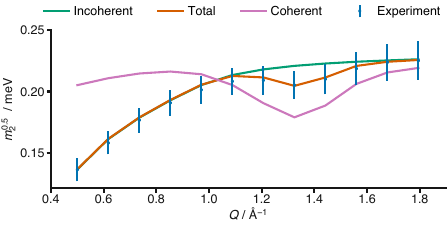}
    \vspace{-1.5\baselineskip}
    \caption{An adaptation of \cref{fig:structure}b to include the coherent contribution showing the characteristic depression of the coherent reduced second moment that is indicative of de Gennes narrowing.}
    \label{fig:si_degenne}
    \vspace{-1\baselineskip}
    \script{si_degenne.py}
\end{figure}

We compare the dynamic structure factor peak integral and reduced second spectral moment of the incoherent and total dynamic structure factor in \cref{fig:structure} of the main text and attribute the discrepancy between these to de Gennes narrowing~\cite{DeGennes1959}. 
Strictly, de Gennes narrowing is the narrowing of the coherent dynamic structure factor at wavevectors corresponding to peaks in the static structure factor. 
To show that what is observed here is indeed de Gennes narrowing, in \cref{fig:si_degenne} we include the coherent contribution in the reduced second spectral moment plot from \cref{fig:structure}b, which shows the depression of the coherent signal, indicative of de Gennes narrowing. 

\section{Translation Diffusion Analysis}
\label{sec:kinisi}

The translational diffusion analysis made use of the \textsc{kinisi} software~\cite{McCluskey2024}. 
The molecular dynamics trajectory was subsampled at every \qty{5}{\pico\second} and the centre-of-mass functionality~\cite{Richardson2025} was used to find the benzene mean-squared displacements at lag times of $\qtyrange{5}{10000}{\pico\second}$ in intervals of \qty{5}{\pico\second}. 
The Bayesian regression scheme implemented in \textsc{kinisi} was then used to estimate the self-diffusion coefficient~\cite{McCluskey2024a}, with the start of the diffusion regime fitting window at \qty{500}{\pico\second}.

\section{Mathematical Derivations of Models}
\label{sec:derive}

\subsection{Rotational Autocorrelation Function}

We start by deriving the rank-$l$ correlation function for an anisotropically axially symmetric rotator. 
Specifically, this considers the oblate symmetric top benzene. 
The Wigner $D$-functions, which form an orthonormal basis set for the 3D rotation group, SO(3), have the following eigenvalues~\cite{Favro1960}, 
\begin{equation}
    \lambda_{lm} = D_tl(l+1) + m^2 (D_s-D_t), 
\end{equation}
where $D_t$ and $D_s$ are the rotational diffusion coefficients for the benzene tumbling and spinning, respectively, $l$ is the magnitude of the angular momentum vector, and $m$ is the projection of that vector~\footnote{These are analogous to the $l$ and $m_l$ orbital angular and magnetic quantum numbers}. 
The rank-$l$ correlation function, $C_l(t, \theta)$ is then defined as, 
\begin{equation}
    C_l(t, \theta) = \sum_{m=-l}^{l} [d^l_{m0}(\theta)] ^ 2 \exp{\left(-\lambda_{lm}t\right)}, 
    \label{eqn:rankl}
\end{equation}
where $d^l_{m0}$ is the appropriate element from the Wigner small-$d$ matrix~\cite{Favro1960, Wigner1964}. 

Using \cref{eqn:rankl}, it is then possible to find any rank correlation function. 
For example, in the main text, we discuss the rotational autocorrelation function, $G_{\theta}(t)$, which follows the rank-\num{2} correlation function, 

\begin{equation}
C_2(t, \theta) = \frac{1}{4} \left(3 \sin^{2}{\left(\theta \right)} - 2\right)^{2} \exp{\left[-t 6D_{t} \right]} + 12 \exp{\left[-t (4 D_{s} + 2 D_{t})\right]} \sin^{4}{\left(\frac{\theta}{2} \right)} \cos^{4}{\left(\frac{\theta}{2} \right)} + \frac{3}{4} \exp{\left[-t (D_{s} + 5 D_{t})\right]} \sin^{2}{\left(2 \theta \right)}.
\end{equation}
This result is equivalent to the Woessner equation used in previous NMR analysis of benzene dynamics~\cite{Woessner1964, Witt2000}. 
From this, the functions that describe the rotational autocorrelation functions for $\theta=\qty{0}{\degree}$ and $\theta=\qty{90}{\degree}$, i.e., parallel and perpendicular to the principal axis of rotation, can be obtained.
These are
\begin{equation}
    C_2(t, \qty{0}{\degree}) \equiv G_{\qty{0}{\degree}} (t) = \exp{\left[-t 6 D_{t}\right]},
    \label{eqn:c0}
\end{equation}
and
\begin{equation}
    C_2\left(t, \qty{90}{\degree}\right) \equiv G_{\qty{90}{\degree}}(t) = \frac{3}{4}\exp{\left[-t\left(4D_s + 2D_t\right)\right]} + \frac{1}{4}\exp{\left(-t6D_t\right)}.
    \label{eqn:c90}
\end{equation}
These can be rationalised, as the vector parallel ($\theta = 0$) to the principal axis is unaffected by the spinning motion, and hence contains only $D_t$. 
Meanwhile, the perpendicular vector is affected by both spinning and tumbling.

\subsection{Dynamic Structure Factor}

Now, we turn our focus to the derivation of the dynamic structure factor. 
We consider a rigid molecule with a single scattering nucleus located at a fixed distance $R$ from the molecular centre. 
Its position in the laboratory frame is, 
\begin{equation}
\mathbf{r}(t) = R\hat{\mathbf{u}}(t).
\end{equation}
where the change in direction of the unit vector $\hat{\mathbf{u}}$ is being probed. 
The intermediate scattering function for incoherent scattering is defined as, 
\begin{equation}
    I(Q, t) = \left\langle \exp{\left(i\mathbf{Q}\cdot\left[\mathbf{r}(t)-\mathbf{r}(0)\right]\right)} \right\rangle,
\end{equation}
where $\langle\ldots\rangle$ denotes an ensemble (and for a liquid, orientational) average. 
Substituting $\mathbf{r}(t) = R\hat{\mathbf{u}}(t)$ and $\mathbf{Q} = \hat{\mathbf{Q}}Q$ into the above gives, 
\begin{equation}
    I(Q, t) = \left\langle \exp{\left(iQR\hat{\mathbf{Q}}\cdot\hat{\mathbf{u}}(t)\right)} \exp{\left(-iQR\hat{\mathbf{Q}}\cdot\hat{\mathbf{u}}(0)\right)} \right\rangle,
\end{equation}
and since the intermediate scattering function is independent of $\hat{\mathbf{Q}}$, we can average over $\hat{\mathbf{Q}}$, 
\begin{equation}
    I(Q, t) = \left\langle\frac{1}{4\pi} \int \exp{\left(iQR\hat{\mathbf{Q}}\cdot\hat{\mathbf{u}}(t)\right)} \exp{\left(-iQR\hat{\mathbf{Q}}\cdot\hat{\mathbf{u}}(0)\right)} \;\textrm{d}\Omega \right\rangle
    \label{eqn:isf}
\end{equation}
where $\Omega$ is the solid angle. 

We can use the standard Rayleigh expansion to reframe these exponential forms using spherical harmonics, i.e., $Y_{lm}$, and spherical Bessel functions, $j_l(x)$,
\begin{equation}
\exp{\left(i\mathbf{k}\cdot\mathbf{r}\right)} = 4\pi \sum_{l=0}^{\infty}i^l j_l(kr) \sum_{m=-l}^{l} Y_{lm}^*(\hat{\mathbf{k}})Y_{lm}(\hat{\mathbf{r}}).
\end{equation}
Therefore, it is possible to rewrite the exponential functions in \cref{eqn:isf} as
\begin{equation}
    \exp{\left(iQR\hat{\mathbf{Q}}\cdot\hat{\mathbf{u}}(t)\right)} =  4\pi \sum_{l=0}^{\infty}i^l j_l(QR) \sum_{m=-l}^l Y_{lm}^*(\hat{\mathbf{Q}})Y_{lm}\left[\hat{\mathbf{u}}(t)\right]
\end{equation}
and
\begin{equation}
    \exp{\left(-iQR\hat{\mathbf{Q}}\cdot\hat{\mathbf{u}}(0)\right)} = 4\pi \sum_{l=0}^{\infty}i^l j_l(QR) \sum_{m=-l}^l Y_{lm}(\hat{\mathbf{Q}})Y_{lm}^*\left[\hat{\mathbf{u}}(0)\right]. 
\end{equation}
At this stage, we observe that we will be able to take advantage of the orthogonality in the spherical harmonics containing $\hat{\mathbf{Q}}$, 
\begin{equation}
    \int Y_{lm}^*(\hat{\mathbf{Q}})Y_{l'm'}(\hat{\mathbf{Q}}) \;\textrm{d}\Omega_{\hat{\mathbf{Q}}} = \delta_{ll'}\delta_{mm'}. 
\end{equation}
Using the orthogonality, we can integrate over $\hat{\mathbf{Q}}$ and use $i^l(-i)=1$ to obtain
\begin{equation}
    I(Q, t) = \left\langle 4\pi
    \sum_{l=0}^{\infty}\left[j_l(QR)\right]^2 \sum_{m=-l}^l Y_{lm}\left[\hat{\mathbf{u}}(t)\right]Y_{lm}^*\left[\hat{\mathbf{u}}(0)\right]
    \right\rangle.
\end{equation}
Next, we use the spherical harmonic addition theorem, which states
\begin{equation}
    \sum_{m=-l}^l Y_{lm}\left(\hat{\mathbf{n}}\right)Y_{lm}^*\left(\hat{\mathbf{n}}'\right) = \frac{2l+1}{4\pi} P_l(\hat{\mathbf{n}}\cdot\hat{\mathbf{n}}'),
\end{equation}
where $P_l$ is the Legende polynomial of order $l$, which is the definition of the rank-$l$ orientational correlation function we derived earlier, i.e., 
\begin{equation}
    C_l(t, \theta) = \left\langle P_l\left[\hat{\mathbf{u}}(t)\cdot\hat{\mathbf{u}}(0)\right]\right\rangle.
\end{equation}
Applying this identity leads to, 
\begin{equation}
    I(Q, t, \theta) = \sum_{l=0}^{\infty}(2l+1) \left[j_l(QR)\right]^2 C_l(t, \theta).
    \label{eqn:isf2}
\end{equation}

From \cref{eqn:isf2}, we can find the intermediate scattering function for an anisotropic axially symmetric rotator, with a maximum $l$ of 2, as, 
\begin{equation}
\begin{aligned}
    I(Q, t, \theta) = & j_{0}(QR)^{2} 
        - 3 j_{1}(QR)^{2} \bigg(\frac{1}{2}\exp{\left[-t (D_{s} + D_{t})\right]} \cos{\left(2 \theta \right)}
        + \frac{1}{2}\exp{\left[-t (D_{s} + D_{t})\right]}
        + \exp{\left(-t 2 D_{t}\right)} \cos^{2}{\left(\theta \right)}\bigg) \\
        & + 5 j_{2}(QR)^{2} \bigg(\frac{1}{4}\left(3 \sin^{2}{\left(\theta \right)} - 2\right)^{2} \exp{\left(-t 6 D_{t}\right)} 
        + 12 \exp{\left[-t (4 D_{s} + 2 D_{t})\right]} \sin^{4}{\left(\frac{\theta}{2} \right)} \cos^{4}{\left(\frac{\theta}{2} \right)} \\
        & + \frac{3}{4} \exp{\left[-t (D_{s} + 5 D_{t})\right]} \sin^{2}{\left(2 \theta \right)}\bigg).
\end{aligned}
\end{equation}
This equation is still dependent on $\theta$, the angle with respect to the principal rotational axis. 
At this stage, we observe that for benzene, all of the scattering nuclei sit in the plane of the molecule, corresponding to $\theta=\frac{\pi}{2}$. 
Therefore, 
\begin{equation}
    I(Q, t) =  j_{0}(QR)^{2} 
        + 3 j_{1}(QR)^{2} \exp{\left[-t (D_{s} + D_{t})\right]}
        + 5 j_{2}(QR)^{2} \bigg(\frac{3}{4} \exp{\left[-t \left(4 D_{s} + 2 D_{t}\right)\right]} + \frac{1}{4}\exp{(-t 6 D_{t})}\bigg).
\end{equation}

From the intermediate scattering functions, the dynamic structure factor can be found; a Fourier transform relates these, i.e.,
\begin{equation}
    S(Q, \omega) = \frac{1}{2\pi} \int_{-\infty}^{\infty} e^{i\omega t}I(Q, t) \;\textrm{d}t.
\end{equation}
For each exponential in the intermediate scattering function, we obtain the following Lorentzian from the Fourier transform, i.e., 
\begin{equation}
    \mathcal{F}\left[\exp{\left(-t\Gamma\right)}\right] = \frac{1}{\pi}\frac{\Gamma}{\omega^2 + \Gamma^2},
\end{equation}
and for $l=0$ we get the elastic line, $j_0^2(QR)\delta(\omega)$, where $\delta(\omega)$ is a Dirac delta-function.
Hence, the total dynamic structure factor is, 
\begin{equation}
    \begin{aligned}
    S(Q, \omega) = \frac{1}{2\pi} \Bigg( & j_0^2(QR) \delta(\omega) + 3j_1^2(QR)\left[\frac{1}{\pi}\frac{D_s+D_t}{\omega^2+(D_s+D_t)^2}\right] \\
     & + 5j_2^2(QR) \left[\frac{1}{4}\frac{1}{\pi}\frac{6D_t}{\omega^2 + (6D_t)^2} + \frac{3}{4} \frac{1}{\pi} \frac{4D_s + 2D_t}{\omega^2 + (4D_s + 2D_t)^2} \right]\Bigg).
    \end{aligned}
\end{equation}
We rewrite this in the form of different $\Gamma_x$ values in \cref{eqn:fit}.

\section{Rotational Autocorrelation Function Analysis}
\label{sec:rotational}

The first \qty{1}{\nano\second} of the production simulation trajectory was taken, and the rotational autocorrelation functions perpendicular and parallel to the principal axis of rotation were found for values of $t = \qtyrange{0}{7.5}{\pico\second}$.
These were expressed with the second-order Legendre polynomial, $P_2$, of the angle $\alpha(t)$, between a molecular axis at time $t$ and its initial orientation at $t=0$, 
\begin{equation}
G_{\theta}(t) = \left\langle P_2 \left[ \cos \left\{\alpha(t)\right\} \right] \right\rangle = \left\langle \frac{1}{2} \left[ 3 \cos^2 \left\{\alpha(t)\right\} -1 \right] \right\rangle,
\label{equ:rot_bes_exp}
\end{equation}
where $\theta$ defines the angle between the vector and the principal axis. 
The resulting rotational autocorrelation functions can be found in \cref{fig:simulation}. 

To extract the tumbling and spinning rotational diffusion coefficients, the data were modelled simultaneously using \cref{eqn:c0} and \cref{eqn:c90}, augmented by additional Gaussian or exponential decay terms to account for dynamics not described by these models.
Bayesian model selection, via nested sampling~\cite{Skilling2004} using the \textsc{dynesty}~\cite{Speagle2020}, was used to determine how many additional Gaussian and exponential decay functions were necessary to fully describe the observed data, with the Bayesian evidence values given in \cref{tab:raf_logz}. 
\begin{table}[h]
    \caption{Log-Bayesian evidence values for different models used to describe the observed rotational autocorrelation function, the model with the highest Bayesian evidence shown in bold (the number of parameters for each model is given in square brackets). In the column where $0$ is added to $G_{\qty{90}{\degree}}(t)$, $A_{\mathcal{N}}$ is a normalisation applied to $G_{\qty{0}{\degree}}(t)$ or $G_{\qty{90}{\degree}}(t)$, and $\mathcal{N}(0, \sigma)$ indicates a Gaussian distribution, centred on $0$ with a standard deviation of $\sigma$.}
    \label{tab:raf_logz}
    \begin{tabular}{ c | c | c}
    \backslashbox{$G_{\qty{0}{\degree}}(t)$}{$G_{\qty{90}{\degree}}(t)$} & $+0$ & $ + (1- A_{\mathcal{N}})\mathcal{N}(0, \sigma)$ \\
    \hline
    $+ (1- A_1)\exp{\left(\sfrac{-t}{\tau_1}\right)}$ & %
  $1103.3\;\;\;[4]$\unskip\unskip\label{output/model_nogauss_oneexp_evidence.txt}\unskip%
 & %
  $1129.7\;\;\;[6]$\unskip\unskip\label{output/model_gauss_oneexp_evidence.txt}\unskip%
 \\ 
    \hline
    $+ A_2\exp{\left(\sfrac{-t}{\tau_1}\right)} + (1 - A_1 - A_2)\mathcal{N}(0, \sigma) $ & %
  $1120.6\;\;\;[6]$\unskip\unskip\label{output/model_nogauss_oneexpgauss_evidence.txt}\unskip%
 & %
  $1140.6\;\;\;[8]$\unskip\unskip\label{output/model_gauss_oneexpgauss_evidence.txt}\unskip%
 \\ 
    \hline
    $+ A_2\exp{\left(\sfrac{-t}{\tau_1}\right)} + (1 - A_1 - A_2)\exp{\left(\sfrac{-t}{\tau_2}\right)} $ & %
  $1124.4\;\;\;[6]$\unskip\unskip\label{output/model_nogauss_twoexp_evidence.txt}\unskip%
 & %
  $\mathbf{1144.3}\;\;\;[8]$\unskip\unskip\label{output/model_gauss_twoexp_evidence.txt}\unskip%
 \\ 
    \hline
    $+ A_2\exp{\left(\sfrac{-t}{\tau_1}\right)} + A_3\exp{\left(\sfrac{-t}{\tau_2}\right)}  + (1 - A_1 - A_2 - A_3)\mathcal{N}(0, \sigma)$ & %
  $1122.6\;\;\;[8]$\unskip\unskip\label{output/model_nogauss_twoexpgauss_evidence.txt}\unskip%
 & %
  $1142.4\;\;\;[10]$\unskip\unskip\label{output/model_gauss_twoexpgauss_evidence.txt}\unskip%
 \\ 
    \end{tabular}
\end{table}

\begin{figure}[h]
    \centering
    \includegraphics[width=\textwidth]{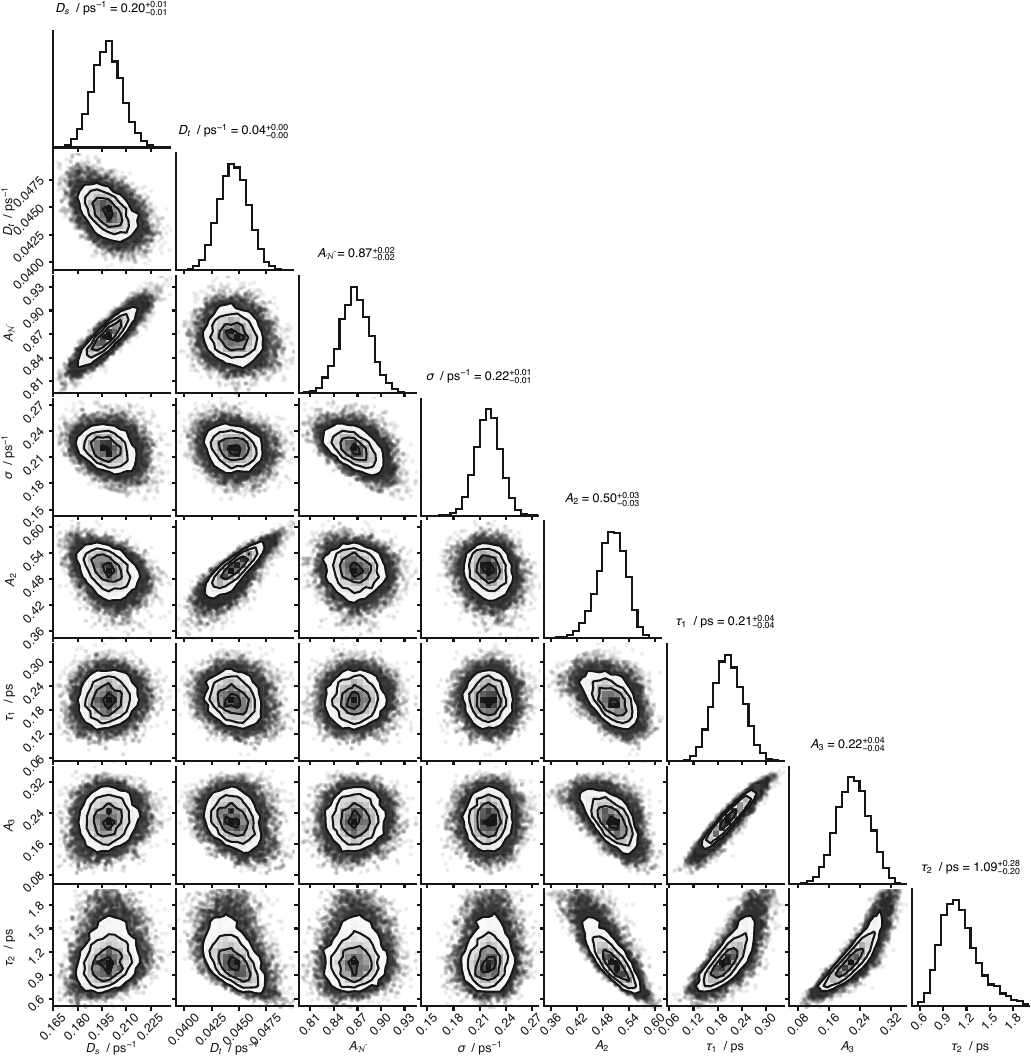}
    \caption{Corner plot showing the posterior distribution of parameters for the highest evidence model given the rotational autocorrelation function data.}
    \label{fig:raf_corner}
    \script{si_raf_chain.py}
\end{figure}

The highest-evidence model consisted of \cref{eqn:c90} with an additional Gaussian distribution to account for very fast dynamics,
\begin{equation}
    G_{\qty{90}{\degree}}(t) = A_n \left[ \frac{3}{4}\exp{\left(-t(4 D_s + 2 D_t)\right)} + \frac{1}{4}\exp{\left(-t6D_t\right)} \right]  + (1- A_{\mathcal{N}})\mathcal{N}(0, \sigma),
\end{equation}
and \cref{eqn:c0} with two additional exponential decays,
\begin{equation}
    G_{\qty{0}{\degree}}(t) = A_1 \exp{\left(-t6D_t\right)} + A_2\exp{\left(\frac{-t}{\tau_1}\right)} + (1 - A_1 - A_2)\exp{\left(\frac{-t}{\tau_2}\right)},
\end{equation}
The full posterior distribution for this model is shown, as a corner plot, in \cref{fig:raf_corner}. 
The fast decay, with time constant $\tau_1 = \qty{0.21}{\pico\second}$, is related to the dramatic decrease observed in the Bayesian evidence at $\qty{1.5}{\milli\electronvolt}$ (corresponding to around $\qtyrange{0.43}{0.52}{\pico\second}$). 
This further justifies the argument made in the main text that it is the presence of fast motions, not present in either analytical model, which leads to the decrease in Bayesian evidence. 

\section{Comparison with Analysis Used in Previous Literature}
\label{sec:previous}

Previous works found a pair of effective characteristic times, $\tau_{\textrm{eff}, \theta}$, by fitting stretched exponential functions to the rotational autocorrelation functions for $\theta=\qty{90}{\degree}$ and $\theta=\qty{0}{\degree}$ and computing the integrals with respect to correlation time, $t$~\cite{Gillen1972, Laaksonen1998, Witt2000, Schwartz2005, Fu2011}.
These integrals can then be used to find $D_{s}$ and $D_{t}$ by solving the system of equations, 
\begin{equation}
    \tau_{\textrm{eff}, \qty{90}{\degree}} = \int_0^\infty G_{\qty{90}{\degree}}(t) \;\textrm{d}t = \frac{3}{4}\frac{1}{4 D_s + 2 D_t} + \frac{1}{4}\frac{1}{6D_t}, 
    \label{eqn:eff90}
\end{equation}
and
\begin{equation}
    \tau_{\textrm{eff}, \qty{0}{\degree}} = \int_0^\infty G_{\qty{0}{\degree}}(t) \;\textrm{d}t = \frac{1}{6D_t}.
    \label{eqn:eff0}
\end{equation}
In practice, this is achieved by fitting a stretched exponential of the form, 
\begin{equation}
    G_{\theta}(t) = \exp{\left[-\left(\frac{t}{\alpha}\right)^\beta\right]},
    \label{eqn:stretch}
\end{equation}
where $\alpha$ and $\beta$ are fitting parameters, the integral is then found as, 
\begin{equation}
    \tau_{\textrm{eff}, \theta} = \frac{\alpha_{\theta}}{\beta_\theta}\Gamma\left(\frac{1}{\beta_\theta}\right).
\end{equation}
This approach introduces a subtle inconsistency: \cref{eqn:c90} describes, for $\theta=\qty{0}{\degree}$, the correlation function as the sum of two exponential functions with different decay rates, whereas the application of the integral at the timescale yields a single effective timescale.
This inconsistency brings doubt into the previously determined values of $D_s$ and $D_t$.  

Applying the approach taken in the literature to our simulation data, gives effective characteristic times of $\tau_{\textrm{eff}, \qty{90}{\degree}} = %
  \qty{1.49}{\pico\second}\unskip\unskip\label{output/lit_approach_t90.txt}\unskip%
$ and $\tau_{\textrm{eff}, \qty{0}{\degree}} = %
  \qty{2.33}{\pico\second}\unskip\unskip\label{output/lit_approach_t0.txt}\unskip%
$. 
Using Eqs.~\ref{eqn:eff90} and~\ref{eqn:eff0}, we can find rotational diffusion coefficients of $D_t=%
  \qty{0.07}{\per\pico\second}\unskip\unskip\label{output/lit_approach_Dt.txt}\unskip%
$ and $D_s = %
  \qty{0.17}{\per\pico\second}\unskip\unskip\label{output/lit_approach_Ds.txt}\unskip%
$. 
These are in good agreement with the literature values presented in previous work~\cite{Laaksonen1998, Witt2000, Schwartz2005, Fu2011}.
However, to our simulation data, we can apply Bayesian model selection to determine whether there is sufficient evidence for the more complex \cref{eqn:exp} over the conceptually simpler stretched-exponential function.
The log-Bayesian evidence for our analytical model for the rotational autocorrelation functions is %
  $1144.3$\unskip\unskip\label{output/best_raf_evidence.txt}\unskip%
, compared to %
  $1052.5\unskip$\unskip\label{output/lit_approach_evidence.txt}\unskip%
 for fitting a stretched exponential function to each rotational autocorrelation function, i.e., there is strong support for the more complex \cref{eqn:exp}. 

\section{QENS Signal from Decomposed Trajectory}
\label{sec:decomp}

To generate the rotational-only simulation trajectory, each benzene molecule was re-centred relative to its molecular centre of mass at $t = \qty{0}{\nano\second}$ for every frame within the full trajectory, with the centre of mass computed via the \emph{pseudo} centre-of-mass recentering method~\cite{Richardson2025}. 
This molecular recentring was applied at each simulation time step in the trajectory, removing all translational motion while preserving molecular rotation.
A complementary translational simulation trajectory was constructed by rigidly fixing the intra-molecular coordinates of each benzene molecule relative to its molecular centre of mass at $t = \qty{0}{\nano\second}$, resulting in a trajectory containing only the translational motion of each molecule.

\begin{figure}[h]
    \centering
    \includegraphics[width=\columnwidth]{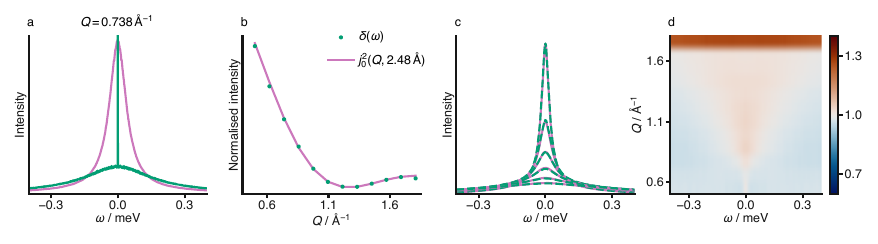}
    \vspace{-1.5\baselineskip}
    \caption{Simulation decomposition: 
        (a) Exemplary rescaled translational (pink) and rotational (green) spectra demonstrating the structural delta peak in the rotational spectra at $Q=\qty{0.738}{\per\angstrom}$, 
        (b) agreement between the decay of the structural $\delta$-function with the zeroth order Bessel function ($R = \qty{2.48}{\angstrom}$)
        (c) Comparison of the first \num{6} simulated incoherent spectra (pink) ($Q_{\textrm{mid}} = \qtylist{0.50;0.62;0.74;0.86;0.98;1.10}{\per\angstrom}$) and the re-convolved decomposed spectra (green dashed).
        (d) Ratio of full simulated incoherent spectra to re-convolved decomposed spectra.}
    \label{fig:si_decomp}
    \vspace{-1\baselineskip}
    \script{si_four_panel_decomp.py}
\end{figure}

%MDANSE settings binning etc
Total and incoherent-only dynamic structure factors for both trajectories were computed using MDANSE~\cite{Goret2017}. 
Example spectra for the rotational and translational components are shown in \cref{fig:si_decomp}a. 
The rotational spectrum exhibits an elastic contribution, represented by a Dirac $\delta$-function, arising from the static spatial correlations in benzene's molecular structure. 
The amplitude of this elastic component follows the decay of the zeroth-order Bessel function evaluated at the molecular radius of benzene ($\sim\qty{2.48}{\angstrom}$) as a function of $Q$, in agreement with theoretical expectations~\cite{Johnson2013}, as illustrated in \cref{fig:si_decomp}b. 
The independent spectra can be re-convolved to reconstruct the incoherent simulated spectra within the dynamical range of interest with minimal loss, as shown in Figs.~\ref{fig:si_decomp}c and \ref{fig:si_decomp}d. 
The only notable discrepancy occurs at very high $Q$, where the re-convolved spectrum reproduces the correct shape but underestimates the amplitude. 
This discrepancy may be attributed to numerical effects, as the signal intensity in this region is extremely low, or unaccounted for roto-translational coupling. 

Models were fit to the rotation-only spectra produced from the decomposed rotational trajectory. 
To determine whether a flat background term was required at each $Q$, the Bayesian evidence was found for a series of 13 models given data, which was binned into the same \num{12} $Q$-bins as the IRIS data, progressively increasing the number of flat background components. 
Background terms were introduced sequentially from the highest to the lowest $Q$.
This ordering reflects the fact that larger $Q$-values probe shorter length scales and therefore faster dynamics, which are more likely to lie outside the instrumental $\omega$-dynamic range of the IRIS spectrometer ($\qty{\pm 0.4}{\milli\electronvolt}$) and manifest as an approximately flat contribution to the spectra. 
For both the isotropic and the anisotropic model, there was the sufficient evidence to fit \num{6} background terms at ($Q_{\textrm{mid}} = \qtylist{1.21; 1.33; 1.45; 1.57; 1.69; 1.81}{\angstrom}$) with log-Bayesian evidences of \num{-99.5} and \num{-71.0} respectively, justifying the additional complexity of the anisotropic model. 
\begin{figure}[h]
    \centering
    \includegraphics[width=0.5\columnwidth]{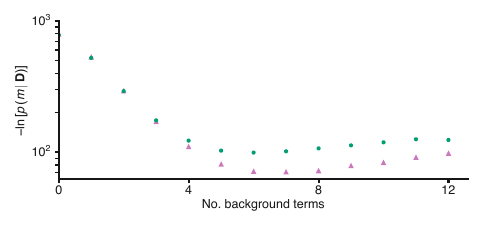}
    \vspace{-1.5\baselineskip}
    \caption{The negative log-Bayesian evidence for the isotropic (green dots) and anisotropic (pink triangles) models for the decomposed rotational QENS spectra as a function of the number of background terms included across $Q$ at \qty{\pm0.5}{\milli\electronvolt} dynamic range.}
    \label{fig:si_rot_nbkg}
    \vspace{-1\baselineskip}
    \script{si_rot_nbkg.py}
\end{figure}

\section{Model fitting}
\label{sec:model_fit}

The full model that was assessed, given the experimental data, was
\begin{equation}
    S_{\textrm{inc}}(Q, \omega) =  C_{\textrm{norm}}\left[\left( e^{-Q^2 \langle u^2 \rangle / 3} \left[\frac{1}{\pi} \frac{D^*Q^2}{\omega^2 + (D^*Q^2)^2} \right] \otimes S_{\mathrm{R}}(Q, \omega) + bkg\right) \otimes S_{\mathrm{res}}(Q,\omega)\right],
    \label{eqn:full_exp}
\end{equation}
where, $bkg$ indicates a flat background describing fast motions beyond the resolution of the instrument, $C_{\textrm{norm}}$ is a normalisation term, $S_{\mathrm{R}}(Q, \omega)$ is either the anisotropic model, \cref{eqn:fit}, or the isotropic model,
\begin{equation}
S_R(Q, \omega) =
j_0^2(QR)\,\delta(\omega) + 3\,j_1^2(QR)
\left[
\frac{1}{\pi}
\frac{2D_r}{\omega^2 + (2D_r)^2}
\right] + 5\,j_2^2(QR)
\left[
\frac{1}{\pi}
\frac{6D_r}{\omega^2 + (6D_r)^2} 
\right],
\label{eqn:iso}
\end{equation}
and $S_{\textrm{res}}$ is the appropriate resolution function data, normalised by the sum of its intensity.

Nested sampling was performed for the model, given the data at incident energies $\qtylist{3.60;1.97}{\milli\electronvolt}$, simultaneously. 
As such, when the nested sampling was performed, the model in \cref{eqn:full_exp} had either 3 or 4 global parameters for the isotropic or anisotropic models, respectively ($D^*$, $\langle u^2 \rangle$, $D_s$, $D_t$, and $D_r = D_t = D_s$ for the isotropic). 
A single $\langle u^2 \rangle$ was used for both incident energies; this parameter describes motions faster than $\omega$-dynamic range and appears as a flat background~\cite{Bee1988}.
$ E_i$-dependent parameters, including a flat background, were included at each $Q$-value for each incident energy in both models. 
A constant normalisation term, $C_{norm}$, was fitted for each incident energy.
$C_{norm}$ was used to normalise across $Q$ for each incident energy, to account for the non-conservation of spectral area introduced by the fast Fourier transform, and to scale to the experimental intensities.
To fulfil the characteristics of a Dirac delta function (integrates to $1$), the delta function intensity at $Q = 0$ was defined as $\delta(\omega)^{-1}$.
For all models, the radius of gyration $R$ was fixed to the radius of the benzene molecule (\qty{2.48}{\angstrom}). 
All model priors were uniform priors of the form  $\theta \in [\theta_{\min}, \theta_{\max})$, where $\theta_{\min}$ and $\theta_{\max}$ for each parameter are given in~\cref{tab:model_priors}.
\begin{table}[h]
\caption{The model prior minimum and maximum values for each parameter.}
\label{tab:model_priors}
\begin{tabular}{l|l|l|l|l|l|l|l}
 & $D^*$ & $D_r$ & $D_t$ & $D_s$ & $\langle u \rangle$ & $C_{norm}$ & $bkg$ \\ \hline
$\theta_{min}$ & 0.01 & 0.01 & 0.001 & 0.06 & 0.001 & 0.01 & 0.01 \\
$\theta_{max}$ & 0.2 & 1 & 0.15 & 1 & 0.18 & 30 & 1.5
\end{tabular}
\end{table}

Due to the distributivity of addition and convolution, the convolution of the rotational and translational Lorentzian components and the Dirac $\delta$-function can be written analytically as a sum of Lorentzian functions. 
In the limit of infinite energy range and continuous resolution, this analytical expression is mathematically identical to a numerical fast Fourier transform-based convolution.
However, in reality, where no analytical resolution function is available, the experimental spectra were measured over a finite energy window and sampled on a discrete energy grid. 
The Lorentzian tails are therefore truncated, and the convolution becomes discrete rather than continuous, breaking the exact equivalence between the analytic and fast Fourier transform convolutions. 
Because the fast Fourier transform convolution naturally incorporates these finite-window and discretisation effects, it provided a representation that more closely matched the measured data. 
For this reason, the fast Fourier transform convolution approach, using \texttt{scipy.signal.fftconvolve}, was used for model fitting.

\section{QENS Model Agreement}
\label{sec:exp_fit}

\begin{figure}[h]
    \centering
    \includegraphics[width=0.5\textwidth]{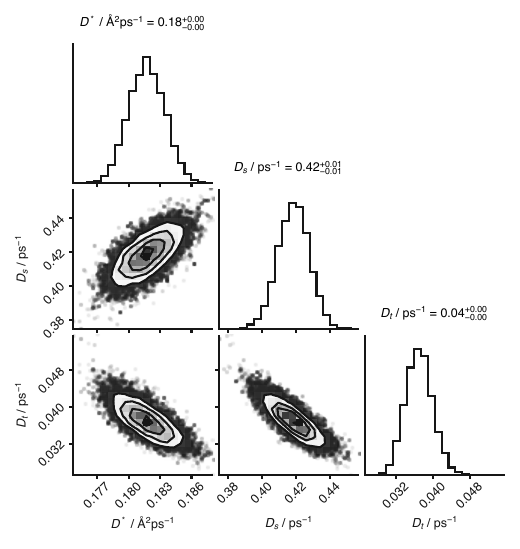}
    \caption{Correlation plot of the Fickian diffusion coefficient, $D^*$ and the spinning, $D_s$ and tumbling, $D_t$, rotational diffusion coefficients from the anisotropic model given the experimental data.}
    \label{fig:si_aniso_corner}
    \script{si_aniso_corner.py}
\end{figure}

\begin{figure}[h]
    \centering
    \includegraphics[width=0.33\columnwidth]{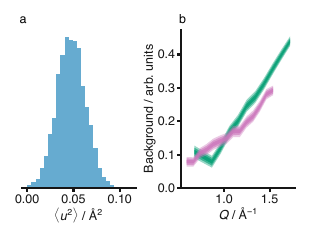}
    \caption{Marginal posterior distributions for (a) the mean-squared displacement (MSD) parameter $\langle u^2 \rangle$, in the Debye–Waller factor in the anisotropic model and (b) the flat background intensity as a function of $Q$ at incident energies of \qty{3.60}{\milli\electronvolt} (green) and \qty{1.97}{\milli\electronvolt} (pink), shading indicates the $1\sigma$, $2\sigma$, and $3\sigma$ credible intervals.}
    \label{fig:si_u_bkg}
    \script{si_u_bkg.py}
\end{figure}

The anisotropic model for the incoherent dynamic structure factor was modelled simultaneously, given the incoherent QENS signal at incident energies of $\qtylist{3.60;1.97}{\milli\electronvolt}$, with a $\omega$-dynamic range \qty{\pm 1.25}{\milli \electronvolt}. 
This modelling yielded the marginal posterior distributions shown in \cref{fig:hist}. 
The correlation corner plot for these marginal posterior distributions is shown in~\cref{fig:si_aniso_corner}; as expected, $D_t$ and $D_s$ are slightly correlated. 
The marginal posterior distributions of the other key parameters, $\langle u^2 \rangle$ and the flat background, as a function of $Q$ for each incident energy are shown in~\cref{fig:si_u_bkg}.

\begin{figure}[h]
    \centering
    \includegraphics[width=\textwidth]{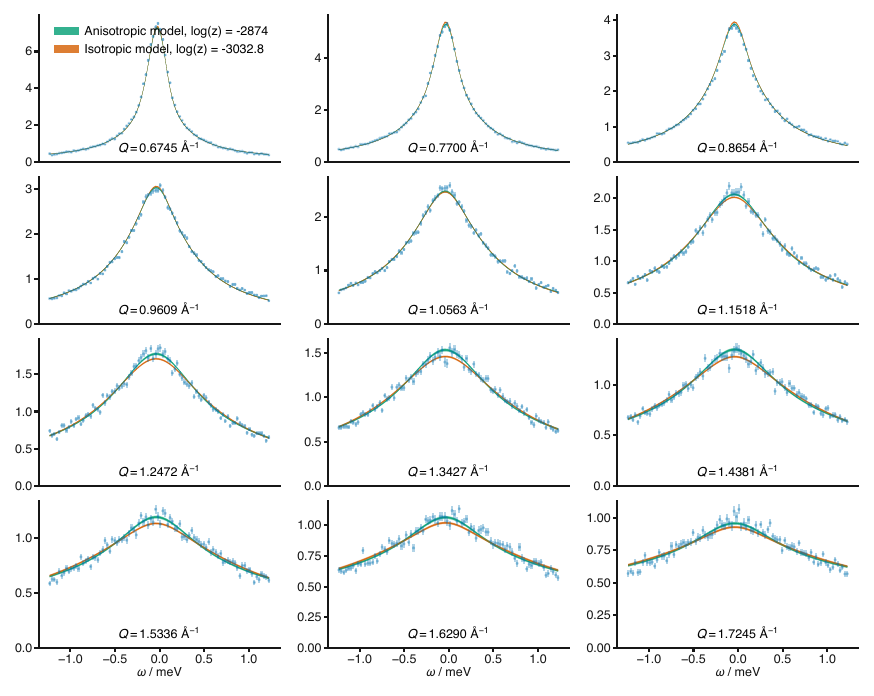}
    \caption{Comparison between the maximum \emph{a posteriori} for the anisotropic model (green) given the binned p-LET data (blue, with errorbars indicating a single standard deviation) at an incident energy of \qty{3.60}{\milli\electronvolt}, at the \num{12} $Q$-bins used.}
    \label{fig:si_model360_fit}
    \script{si_model360_fit.py}
\end{figure}

\begin{figure}[h]
    \centering
    \includegraphics[width=\textwidth]{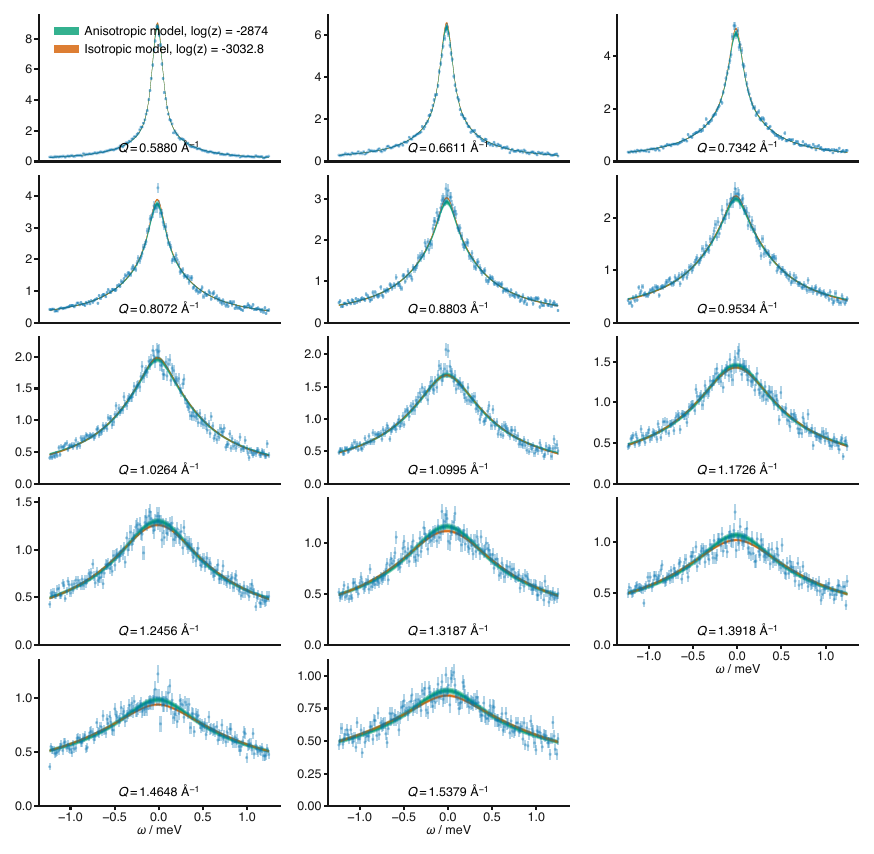}
    \caption{Comparison between the maximum \emph{a posteriori} for the anisotropic model (green) and isotropic model (orange) given the binned p-LET data (blue, with errorbars indicating a single standard deviation) at an incident energy of \qty{1.97}{\milli\electronvolt}, at the \num{14} $Q$-bins used.}
    \label{fig:si_model197_fit}
    \script{si_model197_fit.py}
\end{figure}

The model agreement, represented by the maximum \emph{a posteriori} for each of the incident energies, is shown in \cref{fig:si_model360_fit} (\qty{3.6}{\milli\electronvolt}, 12 Q bins) and \cref{fig:si_model197_fit} (\qty{1.97}{\milli\electronvolt}, 14 Q bins), indicating very good agreement between the model and the data across all $Q$ for both incident energies.
In the interest of clarity, we compare the maximum \emph{a-posteriori} for the isotropic and anisotropic models with the data in Figs.~\ref{fig:si_model360_fit}~\&~\ref{fig:si_model197_fit}. 
There appears to be some structuring present in the ratio between the maximum \emph{a-posteriori} isotropic model and the data (specifically the model underestimates the dynamic structure factor at small $\omega$ and $\qty{1.1}{\per\angstrom} \leq Q < Q_{\textrm{max}}$), which is not observed for the anisotropic model. 
%
% %
% \begin{figure}[h]
%     \centering
%     \includegraphics[width=\textwidth]{figures/si_iso_aniso_compare.pdf}
%     \caption{A comparison of the isotropic and anisotropic model maximum \emph{a posteriori} given experimental LET data at incident energies of $\qtylist{1.97;3.60}{\milli\electronvolt}$ mirroring \cref{fig:qens}.}
%     \label{fig:si_iso_aniso_compare}
%     \script{si_iso_aniso_compare.py}
% \end{figure}
% %

\section{Position and Orientation Probability Density Function}
\label{sec:comp_marta}

\begin{figure}
    \centering
    \includegraphics[width=0.4\columnwidth]{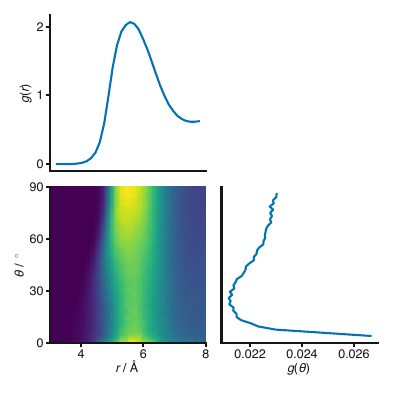}
    \vspace{-1.5\baselineskip}
    \caption{Orientational selectivity by distance showing the centre-to-centre radial distribution function, angular distribution function of molecular pairs and the correlation plot between them.}
    \label{fig:si_angular_density}
    \vspace{-1\baselineskip}
    \script{si_angular_density.py}
\end{figure}

The relative orientations of pairs of benzene molecules were evaluated as a function of their centre-to-centre separation for 2000 randomly chosen frames from the full benzene simulation trajectory (\cref{fig:si_angular_density}).
The intermolecular angle was defined as the angle between vectors parallel to the principal axis of rotation of the benzene molecule, as defined by Falkowska \emph{et al.}~\cite{Falkowska2016}.
The angles were found for molecules between $\qtyrange{3}{8}{\angstrom}$ (in 34 bins) and binned into 45 evenly spaced angular bins between $\qtyrange{0}{90}{\degree}$.
The resulting two–dimensional histogram $g(r,\theta)$ was normalised by the bulk number density and the spherical sector volume corresponding to each $(r,\theta)$ bin,
\begin{equation}
V_{bin} =
\frac{2\pi}{3}
\left(r_{i+1}^{3} - r_i^{3}\right)
\left(\cos\theta_j - \cos\theta_{j+1}\right),
\end{equation}
The radial distribution function $g(r)$ and angular distribution function $g(\theta)$ were obtained by numerically integrating $g(r,\theta)$ over $\theta$ and $r$, respectively.

From the experimental analysis of total neutron scattering data from benzene, Falkowska \emph{et al.} observed the presence of tightly bound ($<\qty{4}{\angstrom}$) $\pi$-stacked benzene dimers and perpendicularly structured dimers in either `T' or `Y' configurations ($\qty{70}{\degree} < \theta \leq \qty{90}{\degree}$, see Fig. 14 in ref.~\cite{Falkowska2016}). 
In our simulations, we observe some presence of the `T' and `Y' configurations, but much less than in the experiment, i.e., no intensity increase at high $\theta$.
Therefore, we do not see evidence of the tightly bound $\pi$-stacked structures. 
We believe that the presence of such dimeric structures may be related to the greater rotational anisotropy observed in our experiment than in the simulation analysis. 

\end{document}